\documentclass[manuscript]{aastex}

\shorttitle{Measuring black hole masses using polarization in lines}
\shortauthors{Afanasiev \& Popovi\'c}

\begin{document}


\title{Polarization in lines - a new method for measuring black hole masses in active galaxies}


\author{Victor L. Afanasiev}
\affil{ASpecial Astrophysical Observatory of the Russian Academy of Sciences, 
Nizhnij Arkhyz, Karachaevo-Cherkesia 369167, Russia}
\email{vafan@sao.ru}

\and

\author{Luka \v C. Popovi\'c}
\affil{Astronomical Observatory Belgrade, Volgina 7, 11060 Belgrade, Serbia}
\affil{Department of Astronomy, Faculty of Mathematics, University of Belgrade, Studentski Trg 16, 11000 Belgrade, Serbia}
\affil{Isaac Newton Institute of Chile, Yugoslavia Branch}
\email{lpopovic@aob.rs}

\begin{abstract}
Measuring of the masses of galactic supermassive black holes (SMBHs) is an important task, since 
they correlate with the host galaxy properties and play an important role in evolution of galaxies. Here we present a new 
method for measuring of SMBH masses using the polarization of the broad lines emitted from active galactic nuclei (AGNs). 
We performed spectropolarometric observations of 9 AGNs and find that this method gives measured masses which are in a good 
agreement with reverberation measurements. 
An advantage of this method is that it can be used to measure the masses of SMBHs in a consistent way at different cosmological epochs. 
\end{abstract}





\section{Introduction}

It is {  widely} accepted that all massive galaxies host supermassive central black holes (SMBHs) ranging in mass from less than a 
million solar masses to many billions of solar masses. The masses of SMBHs are known to correlate with the bulge properties of 
their host galaxies \citep{he11,ko13}, and understanding this is important for understanding the evolution of galaxies 
\citep[see][]{he14}. Therefore,
measuring of the SMBH masses at different cosmological epochs is an important task in astrophysics today.
 There are several methods (direct and indirect) which have been used to measure black hole masses in the center of galaxies 
\citep[see][]{pe14}, where { direct} methods, especially reverberation,
can be used for measuring black hole masses of { low redshift} quasars. 
 However,  measuring of SMBH masses from a single epoch observed spectrum of quasars is still 
 in a developmental stage \citep{pe14}.

 Here  we show how the polarization of the broad lines emitted from  active galactic nuclei 
 (AGNs) can be used to measure the masses of 
supermassive black holes. We tested the method \citep[see][]{af14} for a 
sample of nine AGNs which have been  observed with 6-m telescope of Special Astrophysical
Observatory of Russian Academy of Science (SAO RAS). In difference with reverberation, 
the method only needs one epoch of observation of an AGN.

The paper is organized as following: in \S2 we give method and in \S3 we outline
the basic of observations and obtained results.
 
\section{Method}


We make use of the fact that polarized light in broad AGN lines, can, in some cases, be interpreted as being caused by scattering
in the inner part of a dusty torus \citep[so called equatorial polarization, for more details see][]{sm05,af14}. 
The broad lines are emitted from
the broad line region (BLR) that is close to the SMBH, consequently one can expect near
Keplerian motion of the emission gas in the BLR \citep[][]{ga09}.
A rotating, Keplerian line-emitting BLR surrounded by
a co-planar scattering region produces a polarized broad line, where 
the position angle (PA) of polarization averaged across line profile is aligned with the
projected BLR rotation axis \citep{sm05}, and the stratified velocity field of Keplerian-like motion in the BLR
produces a characteristic change of the PA across the line profile (see Fig. \ref{fig1}, upper panel).  
This gives nearly equal but opposite maximum in the blue and red wings of the line \citep{sm05,af14}.

{ Let us consider a simple model as shown in Fig. \ref{fig1} (panel down), {  considering approximation of a single scattering element 
} from the 
torus. The polarization angle across the line profile, 
$\Delta\varphi(\lambda) = \varphi_L(\lambda)-\varphi_C(\lambda)$\footnote{ where $\varphi_L(\lambda)$
is the PA in the line and $\varphi_C(\lambda)$ is the PA in the continuum}, depends on the velocity field in the BLR \citep{sm05}.}
For Keplerian motion the velocity depends on the distance of the emitting gas from the
SMBH as V$_i\sim R_i^{-1/2}$ and, it also depends on
the polarization angle as V$_i\sim \tan(\Delta\varphi_i)^{-1/2}$ (see Fig.\ref{fig1}, lower panel).  
The relationship between velocities and polarization angles
across the line profile is thus \citep{af14}:
$$\log({V_i\over c})=a-0.5\cdot \log(\tan(\Delta\varphi_i)), \eqno(1)$$
where $c$ is the velocity of light. The constant $a$  directly depends on the black hole mass as
$$a=0.5\log\bigl({{GM_{BH} \cos^2(\theta)}\over{c^2R_{sc}}}\bigr). \eqno(2)$$
where $G$ is the gravitational constant, $R_{sc}$ is the distance of the scattering region from the central black hole
and $\theta$ is the angle between the disc and the plane of the 
equatorial scattering region (see Fig. \ref{fig1}).  Since 
the BLR is expected to be nearly co-planar with the torus, one can take 
$\theta\sim 0$ as a good approximation. {  The effect of a wide (or a non co-planar) torus (e.g. $\theta\sim 10-20^o$) can 
 give an error in black hole mass  (see Eq. (2) $M_{BH}(\theta=0)/M_{BH}(\theta\neq 0)=\cos^2(\theta)$)
measurement around 5-10\%}. There also can be 
partial obscuration of the BLR, but it is connected with the orientation of the system with respect to the line of sight. In principle
one cannot expect a high contribution of the equatorial {  scattering of the line light} in the edge- or face on orientation of the system,
i.e. the equatorial scattering can be expected in the systems with inclination between 20-70 degrees.
On the other hand,  the relation
between velocities  and $\Delta\varphi$ does not depend on the inclination of the system, {  since the Keplerian disc
emits nearly edge-on orientated line light to the scattering region} and relation between 
$\log({V_i})$ and $\log(\tan(\Delta\varphi_i))$ is close to 
0.5 \citep[as it shows a preliminary simulation by STOKES code][]{go14}.

Note here that in  the case where full scattering ring is taken into account
(not a single scattering region as we assumed here), scattered emission will occur at various polarization angles
and there will not be one to one relation between R and $\varphi$. However, in the case of inclined system (torus + Keplerian
disc), the dominant scattered light is coming from one side of the torus, while scattered light for opposite side is blocked, and
the single scattering region approximation can be used \citep[][]{go14}.
More detailed discussion of the model, numerical simulations and exploration of the BLR and scattering region parameters using
polarization in broad lines will be given in an extensive forthcoming paper \citep{af15}.

To measure the black hole mass using Eq. (2) it is necessary to estimate $R_{sc}$.  This is expected to be the inner
radius of the torus \citep{sm05}.  There are two possible ways to find
$R_{sc}$.  The first is to use the empirical relationship between the inner torus radius and luminosity of the AGN \citep[][]{ki11,ko14},
and
the second is to estimate it directly from
infrared observations \citep[][]{ki11,ko14}.

\section{Observations and Results}

To test the method here we have selected a sample of nine AGN (see Table \ref{tab1}),
for which estimates of the inner radius of their tori are given
in literature \citep[][]{ki11,ko14}.  We have obtained spectropolarimetry of the AGN with 6-m telescope of SAO RAS
using a modified version of the SCORPIO spectrograph \citep[for more details, see][]{af11}. The reduction of observed data
and corrections for the interstellar polarization is described in \cite{af14} and here will not be repeated.
The observed AGN and their basic data are given in Table \ref{tab1}. 

The observed shapes of $\Delta\varphi$ and
${V_i\over c}$ for two of the objects, 3C273 and NGC4051, are shown in Fig. \ref{fig2}.  It can be seen in 
Fig. \ref{fig2} that the polarization angle shapes indicate
Keplerian-like motion (Fig. \ref{fig2}, upper panel) and that Eq. (1) fits the observed points relatively 
well except for the low-velocity part (see Fig. \ref{fig2}, lower panel),
where emitting gas is  close to the torus and possibly Keplerian motion is not dominant. { Therefore, we did not considered
points with $V_i\sim 0$, since they are close to the inner radius of the torus and this simple scattering geometry does not work, then
we take $V_{min}$ for  $\Delta\varphi<45^o$}

In Table \ref{tab1} we give estimates of $R_{sc}$ with corresponding literature references 
\citep[columns 4th and 5th, respectively -- see][]{ki11,ko14}, 
estimates for coefficient $a$ (see Eq. (2)), and our black hole masses (columns 6th and  7th, respectively).
We have also compiled reverberation-mapping  estimates of the black hole masses from the 
literature \citep[column 8 in Table 1 -- see][]{pe04,be07,de10,gr12}
and we compare these with our spectropolarimetric estimates (see Fig. \ref{fig3}). As it can be seen in Fig. \ref{fig3}, there is 
a good agreement between our measurements and reverberation ones.  This supports the validity of our proposed method and the good
 correlation between line polarization and reverberation mapping masses strongly supports
our assumption that the scattering region is at or near the inner radius of the torus.

Our method offers a number of advantages over traditional reverberation mapping \citep[see][]{pe14}.  The first and most fundamental 
advantage that is that {\em our estimates of SMBH masses do not depend on the BLR inclination and geometry}, { i.e.
the additional effects to the rotation of the BLR,
as outflows/inflows, which may strongly affect {  the broad line profiles  
(especially widths, which have been used in the reverberation method), do not change  the PA shape across the line profile that is used
in this method. These
effects  can be seen only
as velocity shift of the  PA center (see Fig. \ref{fig2}).}  
A second advantage is that our method only needs one epoch of observation whilst reverberation mapping is 
very telescope-time intensive.  A third advantage is that while in the reverberation method it has to be assumed 
{\it a priori} that the BLR is virialized, in using the spectropolarimetric method one can test the  
assumption Keplerian motion (i.e., virialization) using the relationship between $\Delta\varphi$ and velocities across 
the broad line profile (see Eq. (2) and Fig. \ref{fig2}).  Finally, another significant advantage of our method is that in 
principle it can be applied to lines from 
different spectral ranges going all the way from the near infrared and optical (such as the Balmer lines used as in this paper) to the
UV part (e.g., Ly$\alpha$, CIII], C IV and Mg II) so long as the variation in PA of the polarization can be measured.
This thus allows measurement of black hole masses in a consistent way at different cosmological epochs.
Doing this in practice requires the development of high quality spectopolarimetric instrumentation. 

{ On the other hand, there are some problems with application of the method to some AGNs: i) The
inner torus radius measurements is needed for the black hole mass estimates -- that  can be find by using 
the reverberation method in the infrared \citep[2-3 $\mu$m, see][]{ki11} for low-redshift AGNs, and using the 
calibration between the inner torus radius and the UV radiation for high-redshift quasars \citep[see e.g.][]{ba87}; ii) the method
can be used only for a rotating BLR disk, and in the case of the BLR with dominant radial component (without significant rotation), the 
method cannot be used. It may be a problem with high ionized lines, as C IV and CIII], however, there is indications that also
in high ionized lines a Keplerian motion component is present \citep[see e.g.][]{cl87}. In near future we are going to perform new 
spectopolarimetric observations of high redshift quasars in the UV lines.}



\acknowledgments

Spectropolarimetric observations were obtained with the 6-m BTA telescope of the Special Astrophysical
Observatory Academy of Sciences, operating with the financial support of the Ministry of Education
and Science of Russian Federation (state contracts no. 16.552.11.7028, 16.518.11.7073). The authors also express appreciation to the
Large Telescope Program Committee of the RAS for the possibility of implementing the program
of Spectropolarimetric observations at the BTA.
This work was supported by the  Russian Foundation for Basic Research (grant N12-02-00857)
and the Ministry of Education, Science and Technological Development (Republic of
Serbia) through grant 176001 (``Astrophysical Spectroscopy of
Extragalactic Objects''). 
L.\v C. Popovi\'c thanks the Alexander von Humboldt foundation for supporting his research in general
and COST Action MP1104 ``Polarization as a tool to study the Solar System and beyond''
for supporting his research in this field. We thank Martin Gaskell for useful discussion { and Rene Goosmann for 
giving some explanation with unpublished model simulations. We thank to an anonymous referee for very useful comments.}

\clearpage



\begin{figure}
\epsscale{.80}
\plotone{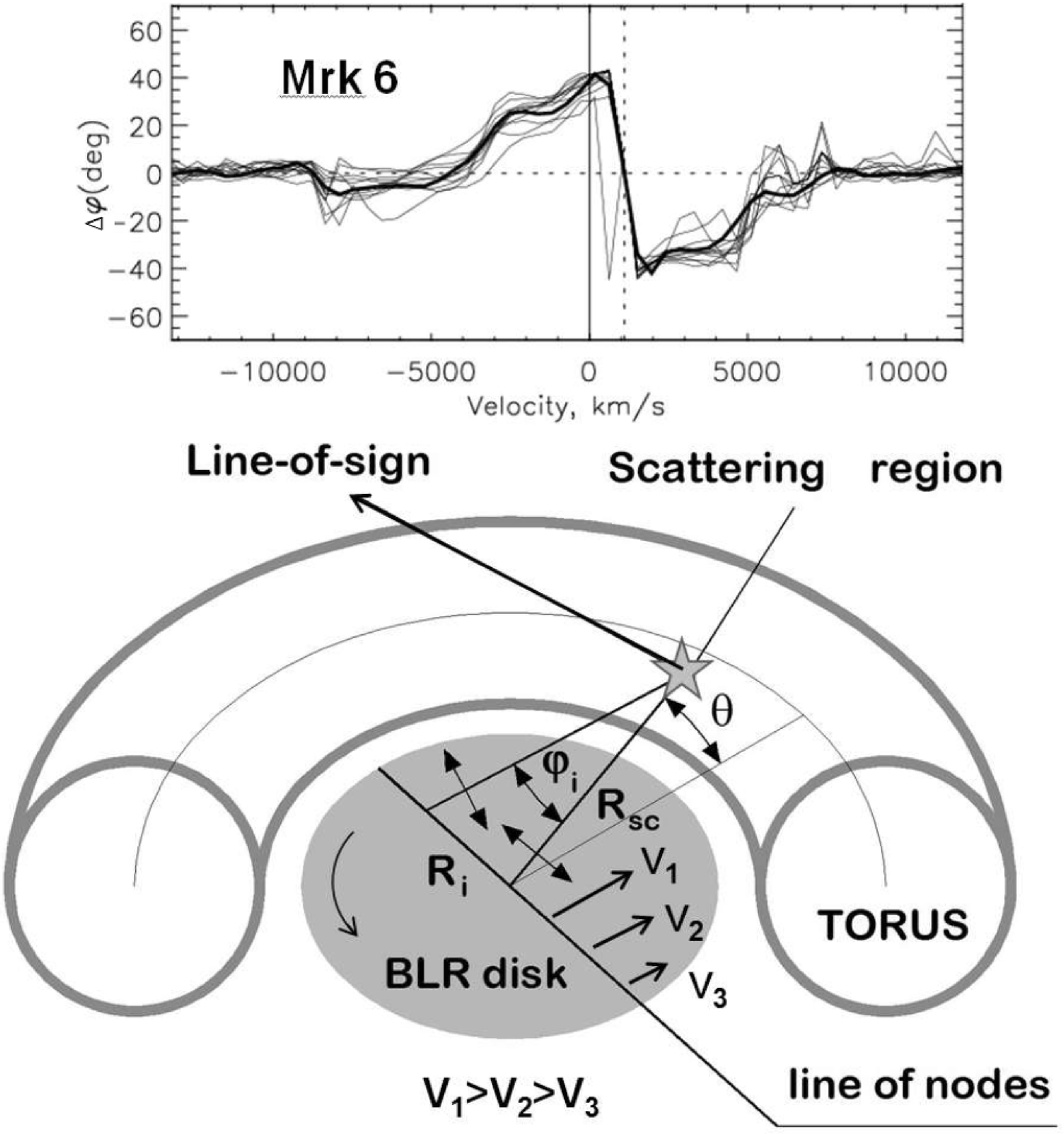}
\caption{Our assumed scattering geometry in indicated in the lower part of the figure and the observed dependence of 
polarization angle ($\Delta\varphi$) vs. 
velocities in the H$\alpha$ line profile of Mrk 6 is shown in the upper part of the figure \citep[see][]{af14}.\label{fig1}}
\end{figure}

\begin{figure}
\plottwo{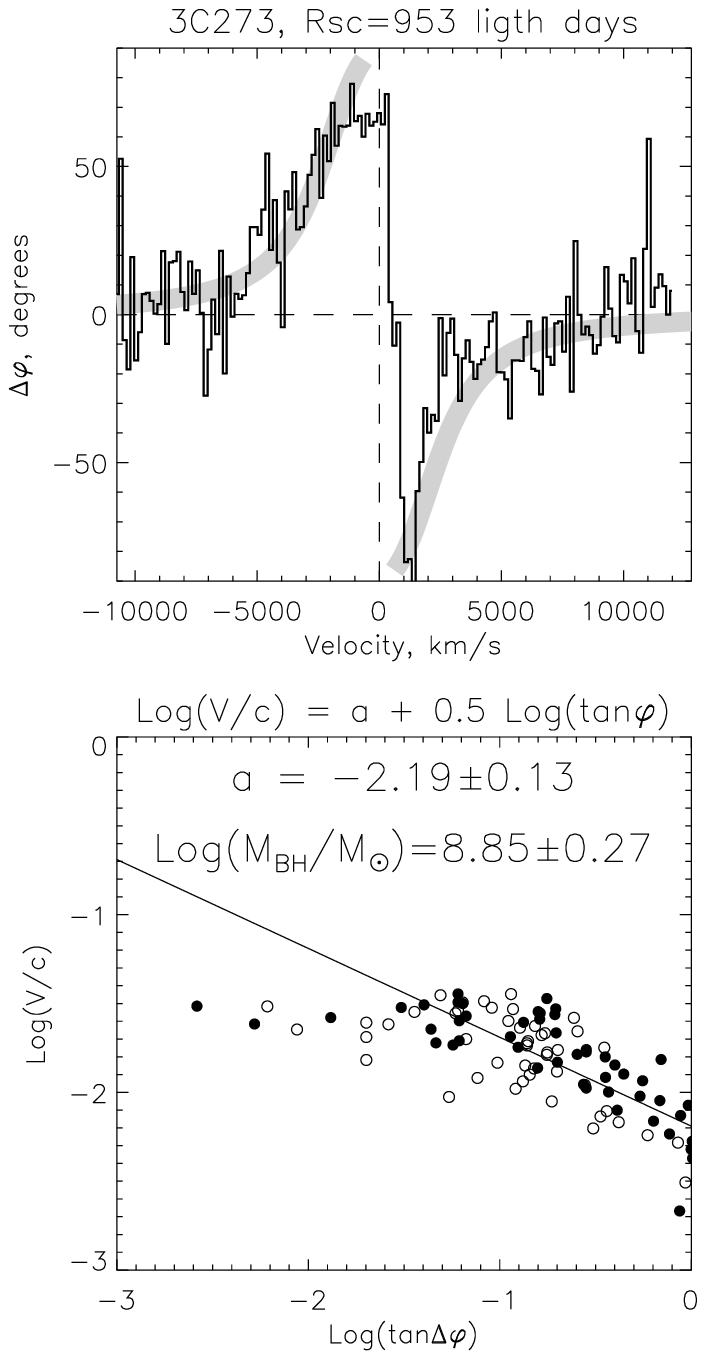}{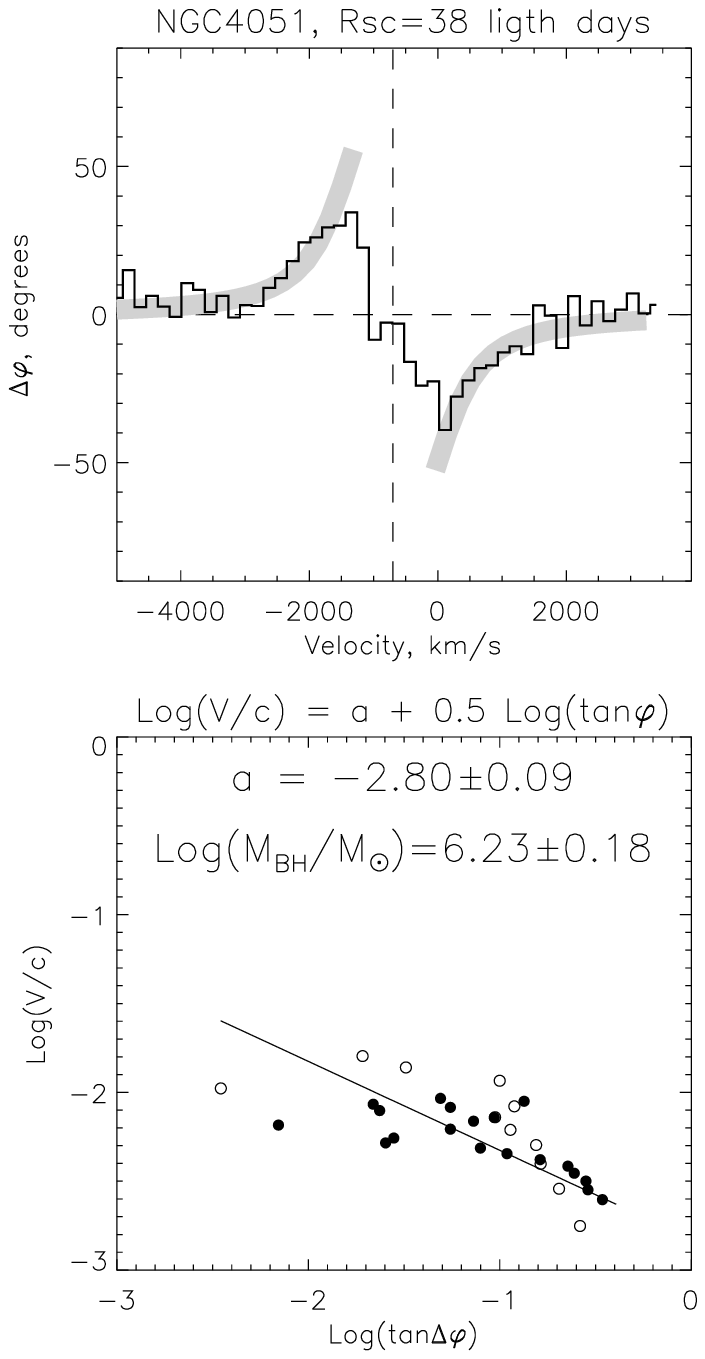}
\caption{The observed shape of  ($\Delta\varphi$) (upper panels) and velocities (lower panels) across the Hα line profile and  for 3C273 
(left hand) and NGC 4051 (right-hand panels) for the AGN in our sample with the highest and lowest masses, respectively. The full
circles are from the blue and open are from the red line part. { The gray curves
in upper panel are expected $\Delta\varphi$ for estimated black hole masses.}\label{fig2}}
\end{figure}

\begin{figure}
\epsscale{.80}
\center
\includegraphics []{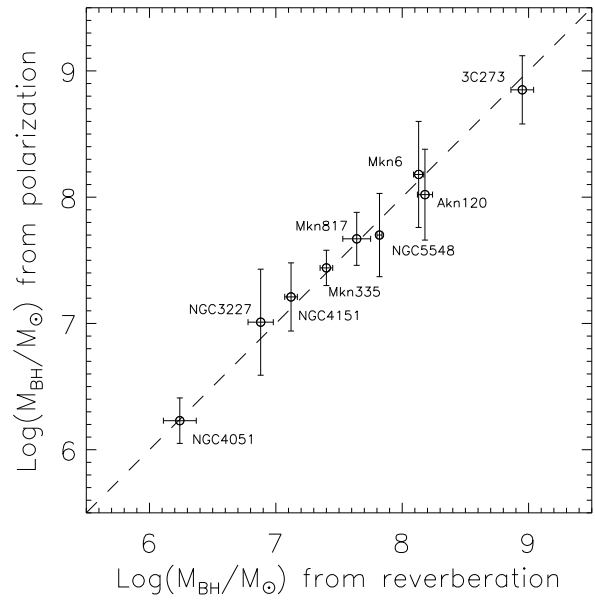}
\caption{Comparison between black hole masses obtained by our spectropolarimetric method with reverberation estimates in the literature
(see Table \ref{tab1}).
\label{fig3}}
\end{figure}

\clearpage
\begin{table*}
\begin{center}
\caption[]{The list of observed AGN with basic data and measured black hole masses using the polarization in broad line (Column 
7)
compared with ones measured by the reverberation method (Column 8). The estimates for R$_{SC}$ are taken from 
1--\cite{ki11} and  2--\cite{ko14}. The reverberation measurements of black hole masses are taken from
 3--\cite{gr12}; 4--\cite{pe04}; 5--\cite{de10} and 6--\cite{be07}}\label{tab1}
\begin{tabular}{lcccccccc}

\hline \hline
Object& Type& z&R$_{SC}$ [pc]&Ref.&$-a$&$\log(M_{POL})[M_\odot]$&$\log(M_{REV})[M_\odot]$&Ref.\\
\hline
Mkn6 &1.5&0.0188&0.185&1&2.19$\pm$0.21&8.18$\pm$0.42&8.13$\pm$0.04&3\\
3C273&1.0&0.1583&0.809&1&2.19$\pm$0.13&8.85$\pm$0.27&8.95$\pm$0.09&4\\
Akn120&1.0&0.0323&0.380&1,2&2.44$\pm$0.18&8.02$\pm$0.36&8.18$\pm$0.06&4\\
NGC4051&1.0&0.0024&0.032&1,2&2.90$\pm$0.09&6.23$\pm$0.18&6.24$\pm$0.13&4\\
NGC4151&1.5&0.0033&0.037&1,2&2.34$\pm$0.13&7.21$\pm$0.27&7.12$\pm$0.05&4\\
Mkn335&1.2&0.0258&0.119&2&2.48$\pm$0.07&7.44$\pm$0.14&7.40$\pm$0.05&3\\
NGC3227&1.5&0.0039&0.021&1,2&2.32$\pm$0.21&7.01$\pm$0.42&6.88$\pm$0.10&5\\
NGC5548&1.5&0.0172&0.096&2&2.30$\pm$0.27&7.70$\pm$0.33&7.82$\pm$0.02&6\\
Mkn817&1.5&0.0315&0.151&2&2.42$\pm$0.11&7.67$\pm$0.21&7.64$\pm$0.11&5\\
\hline
\end{tabular}

\end{center}
\end{table*}
\end{document}